\documentclass[a4paper,twoside]{article}

\usepackage{epsfig}
\usepackage{subcaption}
\usepackage{calc}
\usepackage{amssymb}
\usepackage{amstext}
\usepackage{amsmath}
\usepackage{amsthm}
\usepackage{multicol}
\usepackage{pslatex}
\usepackage{apalike}
\usepackage{url}
\usepackage{algorithm2e}
\usepackage{booktabs}
\usepackage{tabularx}
\usepackage[bottom]{footmisc}
\usepackage{SCITEPRESS} 

\DeclareMathOperator{\FeatureCosineSimilarity}{Feature Cosine Similarity}
\DeclareMathOperator{\FeatureLTwo}{Feature L2 Distance}
\DeclareMathOperator{\FeatureLOne}{Feature L1 Distance}
\DeclareMathOperator{\FeaturePearson}{Feature Pearson Correlation}

\begin{document}

\title{Detecting Concept Drift in Evolving Malware Families Using Rule-Based Classifier Representations}

\author{\authorname{Tomáš Kalný\sup{1}, Martin Jureček \sup{1}\orcidAuthor{0000-0002-6546-8953} and Mark Stamp\sup{2}\orcidAuthor{0000-0002-3803-8368}}
\affiliation{\sup{1}Faculty of Information Technology Czech Technical University, Thakurova 9, Prague, Czech}
\affiliation{\sup{2}Department of Computer Science, San Jose State University, San Jose, CA, USA}
\email{\{kalnyto1, martin.jurecek\}@fit.cvut.cz, mark.stamp@sjsu.edu}
}


\abstract{This work proposes a structural approach to concept drift detection in malware classification using decision tree rulesets. Classifiers are trained across temporal windows on the EMBER2024 dataset, and drift is quantified by comparing extracted rule representations using feature importance, prediction agreement, activation stability, and coverage metrics. These metrics are correlated with both accuracy degradation and data distribution shift as complementary drift indicators. The approach is evaluated across six malware families using fixed-interval and clustering-based windowing in family-vs-benign and family-vs-family settings, and compared against RIPPER and Transcendent baselines. Results show that fixed two-month windowing with feature-level Pearson correlation is the most reliable configuration, being the only one where all family pairs produce positive drift–accuracy correlations. The methods are complementary — no single approach dominates across all pairs.}

\onecolumn \maketitle \normalsize \setcounter{footnote}{0} \vfill

\section{\uppercase{Introduction}}\label{sec1}
Malware evolution remains a major challenge for cybersecurity, as attackers continuously adapt through obfuscation, packing, encryption, and code modification to evade detection \cite{singh_tracking_2012}. This ongoing arms race has reduced the effectiveness of traditional signature- and anomaly-based approaches, leading to widespread adoption of machine learning methods \cite{yan_survey_2023}. Although these methods perform well in controlled settings, their accuracy degrades over time due to concept drift \cite{jordaney_transcend_nodate}.

Concept drift refers to changes in the statistical properties of the target variable over time \cite{lu_learning_2018}, driven by factors such as malware evolution and adversarial obfuscation \cite{jiang_benchmfc_2024}. Without proper handling, drift leads to significant performance degradation, motivating the need for effective detection and adaptation mechanisms \cite{chow_drift_2023}. Existing approaches include statistical monitoring, training with adversarial future samples, and representation learning techniques \cite{lu_learning_2018,jordaney_transcend_nodate,chen_continuous_2023,he_combating_2025,fernando_fesa_2022}.

Rule-based models provide interpretable decision logic and have been successfully applied in malware classification \cite{blount_adaptive_2011,dangelo_association_2021}. However, they remain susceptible to concept drift and require mechanisms to maintain reliability over time. This work addresses drift detection in malware classification through structural analysis of decision tree rulesets, aligning with the growing demand for explainable security solutions \cite{manthena_explainable_2025}.

From an IT security perspective, reliable drift detection is critical for maintaining the effectiveness of deployed systems. In operational environments such as endpoint protection and SOCs, undetected drift can create exploitable blind spots. The proposed approach provides both quantitative indicators of model degradation and interpretable insights into how detection logic evolves, supporting analysts in validation and threat investigation while enabling more adaptive and trustworthy cybersecurity systems.

The work focuses on answering the following questions:
\begin{itemize}
    \item Can changes in decision tree rules be used as meaningful drift indicators?
    \item Do such drift indicators correlate with performance degradation or data distribution
shifts?
    \item How does drift detection differ between malware vs. benign and malware family vs. malware
family classification settings?
    \item  What is the minimal temporal resolution is required to observe meaningful drift signals?
\end{itemize}

The remainder of this paper is organized as follows. Section~\ref{sec:related_work} reviews related work, Section~\ref{sec2} provides background on concept drift in malware, Section~\ref{sec:proposed_method} describes the proposed method, Section~\ref{sec:exp_results} presents experimental results, and Section~\ref{sec:conclusion} concludes the paper.

\section{\uppercase{Related Work}}\label{sec:related_work}

Jordaney et al.\ introduced the Transcend framework \cite{jordaney_transcend_nodate}, which detects out-of-distribution samples using conformal evaluation. Rather than relying on classifier confidence scores, Transcend computes independent p-values per class via non-conformity measures, making it model-agnostic and less prone to overconfident predictions on drifted samples. The framework was later extended by Barbero et al.\ \cite{barbero2022transcending} with improved calibration. We compare against Transcend directly in Section~\ref{sec:exp_results}.

Chow et al. \cite{chow_drift_2023} investigated the root causes of concept drift in Android malware samples using the same data set as Transcend framework. Their methodology involved identifying points of interest where drift occurred by comparing the predictions of a base model against oracle models trained on subsets of future test data. This enabled them to determine which malware families contributed the most significantly to drift and to assess the corresponding impact on classifier performance.  Interestingly, the results showed that benign samples are not the main source of drift and have very little impact on performance. 

Dolejš et al. \cite{j_dolejs_interpretability_2021} explored the explainability and interpretability of ML-based malware detection systems. Their study focused on Windows PE files from the EMBER dataset and employed rule-based models I-REP and RIPPER for better explainability and their ability to handle noisy data. To evaluate interpretability, the authors proposed two metrics: Human Most Understandable Model and Interpretability Entropy. In their experiments, five machine learning algorithms were first trained on static features; their predictions were then used to train rule-based models, reconstructing the decision making process through rule sets. The results demonstrated that the best performing combination was to use RIPPER on Gaussian Naive Bayes (GNB) predictions, achieving an accuracy of 98.49 percent covering most of the predictions. This highlights the potential of rule-based post-processing to provide transparency in otherwise opaque classifiers.

Jurečková et al. used rule-based classifier RIPPER to detect concept drift in 6 malware families using EMBER dataset and MAB-malware adversarial generator to simulate drift and get a ground-truth moment when drift has occurred. Applying the same rule-based classifier allowed authors to directly compare features of original and evolved samples based on conditions that the samples meet and rule set distance function. Authors tested multiple feature dimensionalities from 3 to 100 and reached overall 92.08\% accuracy of detecting drift across all 6 families. \cite{jureckova2026detecting}. 

Stamp et al. \cite{mishra2025cluster} propose a lightweight clustering-based approach for concept drift detection in setting that does not require class labels. They apply MiniBatch K-Means to temporally-ordered batched from KronoDroid dataset and track changes in silhouette coefficient between consecutive batch pairs. Their approach has been validated using three scenarios - static retraining (baseline), periodic retraining and drift-aware retraining across four classifiers. With drift-aware retraining the authors were able to reduce the number of models trained by approximately \(40\%\) while achieving accuracy within 0.5\% of periodic retraining.

Ceschin et al. proposed updating both the classifier and the feature selector when there is a certain level of drift \cite{ceschin_fast_2023}. The authors studied whether drift remains across datasets. Compared to just retraining the classifier, the proposed "Fast and Furious" solution achieved an improvement of 22.05\% and 8.77\% for the F1 score in the Drebin and AndroZoo datasets, respectively. 

Other works such as FeSA \cite{fernando_fesa_2022} study the creation of an optimal feature selection architecture for ransomware classification when exposed to concept drift. The multilayered architecture for feature selection is proposed in order to decrease classifier degradation. Singh et al. \cite{singh_tracking_2012} focus on tracking concept drift within malware families using relative temporal similarity and meta-features. Interestingly, their results suggest that there is no significant drift within the malware families they studied; however, the authors also mention that the number of families they studied is relatively small and they have only looked at static features . Lastly, Darem et al. \cite{darem_adaptive_2021} proposed incremental adaptive batch learning to continuously retrain the model. In order to monitor drift, they monitor error statistics of the classifier. It is worth noting that several of the discussed approaches require timely access to accurate class labels, which remains a practical challenge in malware detection where ground truth is often obtained with significant delay.

\section{\uppercase{Concept Drift}}\label{sec2}
Under concept drift the statistical properties of the training and testing data distributions diverge over time, undermining the basic principle of machine learning, which assumes that the training and test data come from the same probability distribution. The main sources of drift in malware are changes made to malware families themselves, such as new functionality, obfuscation techniques, or adaptations to different environments, which manifest as changes in the conditional distribution of the target variable or shifts in the input data distribution \cite{singh_tracking_2012}. Due to concept drift, a detection system that has been trained on patterns from the past may not be able to recognize newly evolved malware that contains unseen patterns. This leads to outdated, ineffective, and at its core unusable models.

Lu et al.\ \cite{lu_learning_2018} define four types of concept drift (sudden, gradual, incremental, reoccurring) and categorize detection algorithms into error-rate-based, data-distribution-based, and multiple hypothesis test methods. Error-rate-based methods, while the most common, are of limited applicability in malware detection as they require immediate access to true labels, which are typically obtained with significant delay.

The work \cite{singh_tracking_2012} provides a simpler view for handling concept drift, namely direct approaches which regularly adapt the classifier without detecting concept drift, and so-called detect and adapt approaches, which monitor the level of drift using features of data, parameters of the classifier, and performance metrics. 

\subsection{Concept Drift in Malware}
Malware families evolve through various mechanisms.
Sing et al. \cite{singh_tracking_2012} identify three main categories; natural evolution resulting from fixing the code/adding new functionalities to the malware/ creating versions for other environments. Changes in software in which the malware was developed cause environmental evolution; these include different compiler versions, library updates, etc. And polymorphic evolution, the most relevant to our work, occurs when malicious attackers use evasion techniques such as packing, encryption, adversarial attacks, etc. to avoid malware detectors. 

Jiang et al. \cite{jiang_benchmfc_2024} differentiate between two causes of malware drift, a semantic shift caused by introducing new, previously unseen malware families, and a covariate shift caused by a change in the input distribution (packed or evolved families).

Various strategies have been proposed to mitigate the effect of concept drift. Most works use retraining of the model for which multiple approaches exist, such as retraining periodically \cite{mishra2025cluster}, retraining when concept drift occurs with new data, and retraining using dynamic batches of both new and old data so that the model does not forget old patterns. 

The methods for detecting drift and selecting the appropriate samples for retraining vary considerably between studies \cite{he_combating_2025,darem_adaptive_2021}. Ensemble learning combats concept drift by training multiple classifiers from different time windows and classifying the sample with a weighted combination of outputs from these models. An alternative direction is to use adversarial training, in which adversarial samples are introduced during training in hopes of making the model more robust for future unknown samples \cite{bosansky_counteracting_2024}. 

\section{\uppercase{The proposed method}}\label{sec:proposed_method}
In this section, we describe the experimental design used to evaluate our multi-metric structural drift detection approach. We evaluate the following experimental scenarios:
\begin{itemize}
    \item varying window sizes,
    \item binary decision trees trained on a malware family versus benign samples compared to decision trees trained on pairs of closely related malware families,
    \item per-family clustering to identify malware campaigns as an alternative to fixed time windowing.

\end{itemize}

\subsection{Preprocessing of data and feature selection}
Duplicate samples were removed using SHA-256 hashes. Timestamps were extracted from COFF headers of PE files and approximately 10\% of samples were excluded due to inconsistencies with VirusTotal first-submission dates (e.g., COFF timestamp newer than submission, or differences exceeding the 95th percentile). While removing these samples may slightly simplify the classification task, retaining unreliable timestamps would compromise the temporal windowing fundamental to our drift detection approach.

For our experiments, we have used EMBER dataset \cite{joyce2025ember}, which is high-dimensional (2568 features), leading to increased model complexity, longer training times, and reduced interpretability of decision trees. To address these issues, we applied feature selection prior to model training. Specifically, we used the SelectKBest method from the scikit-learn library\footnote{https://scikit-learn.org/stable/} \cite{scikit-learn} which ranks features by a univariate statistical test and retains the k
k highest-scoring features.  Evaluated feature subset sizes ranged from 1 to 20. The final subset of 10 selected features is listed in Table~\ref{tab:selected_features}. Our hypothesis for the dramatic reduction from 2568 to 10 features while still maintaining accuracy can be explained by the nature of the EMBER feature set, which encodes a broad structural profile of PE files. For the task of distinguishing specific malware families, only a small subset of these structural characteristics carries discriminative information, as individual families tend to differ in specific features such as entropy patterns or import profiles rather than across the entire feature space. 

\begin{table}[ht] 
\centering 
\caption{Selected features after SelectKBest feature selection.} 
\label{tab:selected_features} 
\begin{tabularx}{\linewidth}{@{}l X@{}} 
\toprule 
Category & Feature \\ 
\midrule 
Byte entropy & \texttt{bin\_116}, \texttt{bin\_117} \\ 
Section entropy & \texttt{entropy\_hash\_19} \\ 
Section characteristics & \texttt{char\_hash\_1}, \texttt{char\_hash\_3}, \texttt{char\_hash\_13}, \texttt{char\_hash\_35} \\ 
Imports & \texttt{lib\_hash\_3}, \texttt{lib\_hash\_52} \\ 
PE file warnings & \texttt{warn\_67} \\ 
\bottomrule 
\end{tabularx} 
\end{table}

During our experiments, the classification accuracy plateaued at 10 features; we fixed this value for all experiments (Table~\ref{tab:selected_features}).

\subsection{Evaluation process}\label{subsec:drift_metrics}
This section contains a detailed explanation of how we evaluated our experiments and what drift and data shift metrics were defined in the process. 
\subsubsection{Drift metrics}
We proposed and evaluated multiple drift metrics to quantify changes between family evolutions over time windows. First, let us provide a notation used in formulas for metrics. Let $R_t$ denote the ruleset extracted at time $t$, where each rule $r \in R_t$ is a conjunction of conditions of the form $f \odot h$ (feature, operator, threshold). We denote the dataset as $X$, the number of features as $N$, the feature importance vector as $v^{(t)}$, and rule coverage as $\mathrm{Coverage}(r, X) = |\{x \in X : x \models r\}| / |X|$.
The metrics are defined as follows:
\begin{itemize}
    \item Feature importance drift - These metrics were devised by us and are based on feature importance vector derived from the extracted ruleset of decision tree. Feature importance is computed, based on how frequent each feature appears in the rule sets, weighted by the length of the rules seen in Eq. \eqref{eq:feat_importance_score}. Given two rule sets from different time windows, we compute distances between their corresponding normalized feature importance vectors using cosine similarity (sensitive to directional change) noted as $\FeatureCosineSimilarity$, Euclidean distance ($\FeatureLTwo$) (magnitude of importance change), Manhattan distance (measuring total amount of change, noted as $\FeatureLOne$) or Pearson correlation (capturing changes in relative feature ranking) noted as $\FeaturePearson$. Feature importance is computed as 
\begin{equation}
s_i=\sum_{r\in R}\frac{|\{c\in r: feat(c)=f_i)\}|}{|r|}\label{eq:feat_importance_score}
\end{equation} 

where \(|\{c \in r: feat(c)=f_i)\}|\)counts how many conditions in rule r reference feature \(f_i\).

The normalized feature importance vector (where \(\sum v_i = 1\) is computed as
\begin{equation}
    v_i=\frac{s_i}{\sum^N_{j=1}s_j}
\end{equation}

\item Prediction agreement --- The fraction of samples where both rule sets produce the same prediction, defined as
    \begin{equation}
\text{Agreement}=\frac{1}{|X|}\sum_{x \in X}\mathbf{1}\!\left( y_{R_{t-1}}(x) = y_{R_t}(x) \right),
\end{equation}
where $y_R(x)$ is the binary prediction of ruleset $R$ on sample $x$.
\end{itemize}
Additionally, we evaluate two auxiliary metrics: \emph{coverage mean diff windows}, measuring the change in average rule coverage between consecutive windows, and \emph{activation stability drift}, defined as the cosine similarity of per-rule activation rate vectors across windows.

These metrics are correlated against two ground-truth hypotheses about when drift has occurred. The first hypothesis assumes that drift manifests as an accuracy drop in the subsequent time window; accordingly, we calculate the correlation between the drift metric in window $t$ and the accuracy change in window $t+1$. The second hypothesis uses data distribution shift as an indicator of drift, correlating the drift metric in window $t$ with the distribution shift observed in window $t+1$. For each test configuration, we first report mean correlations for the strongest metrics across all windows, then select the two strongest metrics based on these mean correlations and examine their per-family results.

We use Spearman rank correlation as it captures monotonic (not only linear) relationships, is robust to outliers given our small sample of windows, and is invariant to the heterogeneous scales of our metrics. 

\subsubsection{Data distribution shift}
As discussed earlier, data distribution shift is treated as the ground-truth indicator of drift in our experiments. To quantify this shift, we defined several data distribution shift metrics:
\begin{itemize}
    \item Mean L2 \_Distance: Geometric metric, which measures the distance between two mean feature vectors of two consecutive windows. 
    \item Domain classifier AUC: A multivariate metric, that captures whether the relationships between features have changed. For a given family, samples from the previous window are labeled as class 0 and samples from the current window as class 1. A Random Forest classifier is trained to distinguish between them, and the  Receiver Operating Characteristic (ROC) and Area Under Curve (AUC) is computed. An AUC close to 0.5 indicates highly similar distributions, while higher values indicate stronger distributional shifts.
    \item Mean Wasserstein: A univariate metric computed per feature and then averaged. For each feature, we calculate the Wasserstein distance using scipy library \cite{scipy} (work required to transform the distribution of previous window into distribution of the current window). Then average of these distances is computed. Values close to zero indicate minimal distributional change, while larger values suggest increasing drift. 
    \item KS mean: An aggregation of per-feature Kolmogorov-Smirnov tests \cite{scipy}. The score ranges from 0.0 to 1.0, if distributions are identical the KS mean is 0, 1.0 for distributions with no overlap. 
\end{itemize}

\subsubsection{Fixed windowing}
The evaluation process consists of the division of training and testing data into windows based on the timestamp and the fixed range of the time window. If retraining is enabled, the classifier is retrained in each time window. Specifically, for a given window \(t\), the classifier is trained using the training data from the window \(t\) and evaluated on the testing data from the subsequent window \(t+1\).

To increase the number of evaluation samples available for drift metric computation, we concatenate training and testing data from the time window $t+1$; however, training is performed exclusively in window $t$. This concatenation improves the reliability of coverage-based metrics such as Jaccard Cover rules and Prediction Agreement, which can become noisy when computed on small test sets alone, particularly in short windowing configurations where some families may have very few samples (sometimes even 0 when windowing by weeks). Note that this introduces a size asymmetry between the sample sets from windows $t$ and $t+1$; however, the affected metrics are computed as normalized ratios (e.g., fraction of covered samples, fraction of agreeing predictions), which mitigates the impact of differing sample sizes. During the evaluation, we compute the accuracy, all defined drift metrics, and data distribution shift metrics.

The primary objective of this evaluation is to determine whether drift metrics derived from decision tree rulesets can predict concept drift. Since we do not know exactly when the drift has occurred for the family, we define two complementary scenarios.

In the first scenario, we assume that the concept drift causes a decrease in the accuracy of the model. This assumption is based on the fact that adversaries modify malware samples in order to avoid detection, meaning intentional reduction of accuracy. Therefore, we calculate correlations between the accuracy difference in the next window with drift metrics from time window t. We expect the drift to have a negative impact on future predictions. 

In the second scenario, we treat the data distribution shift as a ground-truth drift. We measure the per-family distribution shift between consecutive windows  using multiple data shift metrics. These metrics are then used as baseline drift indicators, and we compute the correlations between them and the ruleset-based drift metrics.

\subsubsection{Clustering}
In the cluster-based evaluation, the overall pipeline remains identical to the fixed windowing, except for the temporal segmentation of data. Instead of using fixed-length time windows, we identify per-family malware campaigns by clustering samples in time, grouping periods of family activity. Two clustering algorithms were evaluated, specifically DBSCAN and KMeans.

Once family-specific clusters are identified based on pairs of families, each cluster is treated analogously to a time window, and the subsequent training, evaluation, and drift analysis steps remain unchanged.

\section{\uppercase{Experimental Design and Results}}\label{sec:exp_results}
This section presents the design and results of our experimental evaluation of the proposed drift metrics. Our focus is on the relationship between decision tree ruleset changes and classifier accuracy degradation and data distribution shifts which are our ground-truth drift baselines.

\subsection{Experimental Design}\label{sec:experimental_setup}
All experiments were conducted on the EMBER2024 dataset \cite{joyce2025ember}, specifically using the PE file subset. The dataset was restricted to the six largest malware families (Table \ref{tab:family_counts}). This choice was motivated by several factors: reducing computational cost, enabling more reliable per-family analysis, and ensuring that each time window contained a sufficient number of samples for all families.

\begin{table}[htbp]
\centering
\caption{Number of samples per malware family}
\label{tab:family_counts}
\begin{tabular}{lr}
\toprule
\textbf{Family} & \textbf{Count} \\
\midrule
berbew  & 174{,}466 \\
wacatac & 77{,}727  \\
expiro  & 74{,}339  \\
cosmu   & 53{,}965  \\
xmrig   & 28{,}633  \\
upatre  & 25{,}296  \\
\bottomrule
\end{tabular}
\end{table}

To limit overfitting, control model complexity, and ensure comparability across experiments, we set the decision tree hyperparameters throughout all experiments to the same values. The maximum tree depth was set to 6, the random seed to 52, the splitting criterion to Gini impurity, the minimum number of samples per leaf to 5, and the number of features considered at each split was not restricted.

The implementation was written in Python 3.10 and used primarily libraries such as Polars\footnote{https://pola.rs} for operations with dataframes, Scikit-learn\footnote{https://scikit-learn.org/stable/} for all the machine learning, Numpy\footnote{https://numpy.org} and Scipy\footnote{https://scipy.org}.

\subsection{Fixed windowing}
This subsection presents results obtained using fixed-size temporal windows in both family-versus-benign (FvB) and family-versus-family (FvF) classification settings, as well as under different temporal resolutions.

\subsubsection{Family-versus-benign results}
In the FvB setting with 2-month windows and periodic retraining, accuracy stabilized around 98\% for xmrig, berbew, expiro, and cosmu, while wacatac dropped to 60\% mid-year. Aggregated correlations between drift metrics and next-window accuracy difference remained modest, with feature importance-based metrics clustering around mean correlations of 0.2 and exhibiting high standard deviation. Per-family analysis revealed strongly family-dependent behavior: for example, Feature L1 distance correlated at $0.8$ for wacatac but $-0.85$ for cosmu at lag $k{=}1$, while Feature Cosine similarity showed mostly opposite signs. These results indicate that no single drift metric provides reliable, family-independent prediction of accuracy degradation in the FvB setting. Restricting the drift computation population to only family and benign samples had negligible effect on the results.

Shortening the window to 2 weeks with periodic retraining yielded substantially weaker correlations (highest aggregated mean $\rho = 0.109$), confirming that concept drift in malware families accumulates gradually and is best captured on broader temporal scales.

Data distribution shift correlations in the FvB setting yielded comparable values to the accuracy-based analysis. The strongest aggregated correlations reached approximately $0.25$ for combinations involving feature importance metrics and KS mean or Mean Wasserstein, again with high variability across families.

\subsubsection{Family-versus-family accuracy correlation}
In the FvF setting, the binary classifier was trained to distinguish between pairs of malware families, selected based on medoid distances. Rule extraction followed the same procedure as in the FvB setting; although the classifier is trained on two families, each ruleset is associated with a single family by treating the second family as the negative class.

\begin{figure}
    \centering
    \includegraphics[width=1\linewidth]{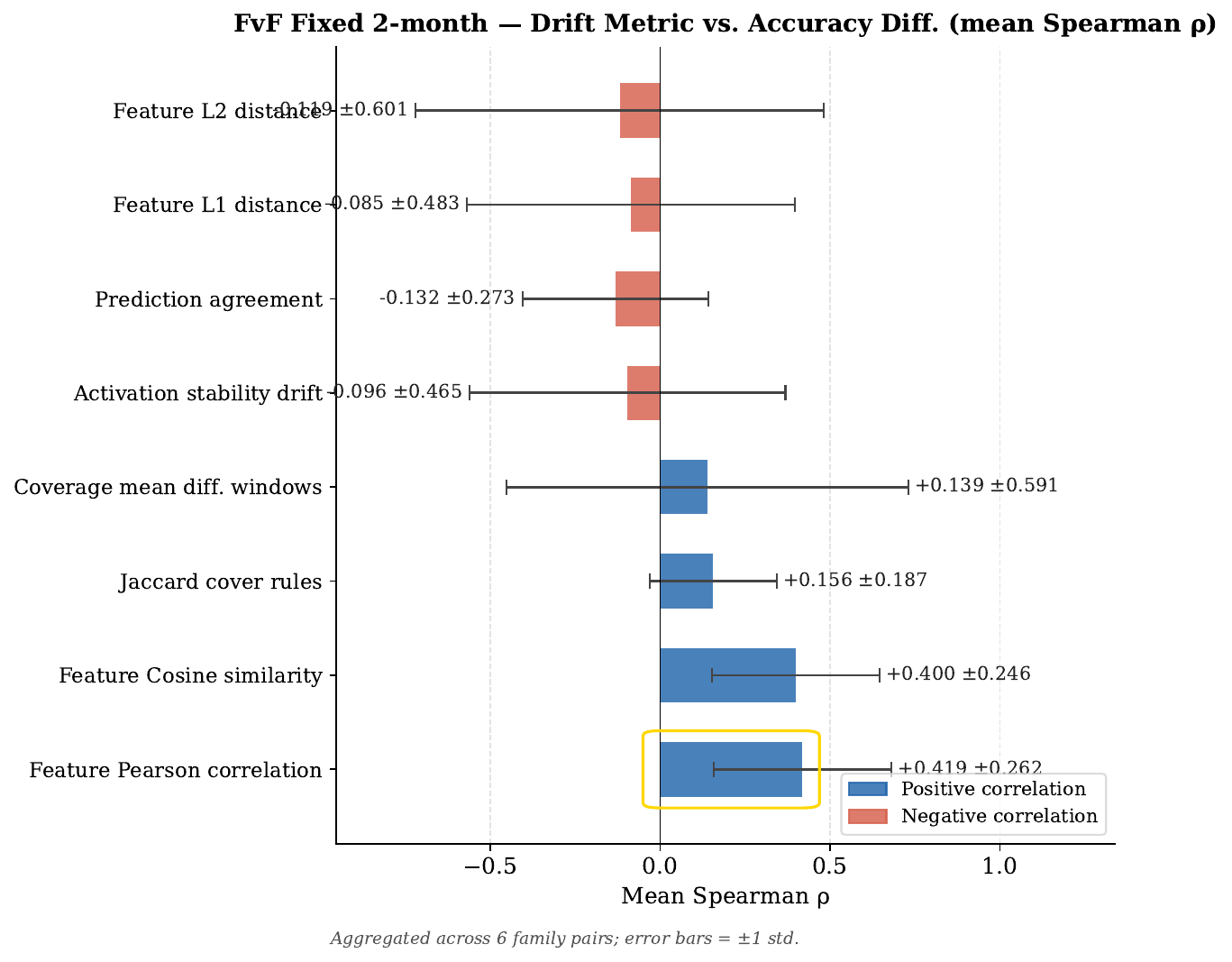}
    \caption{Top correlations (mean across families) for family-versus-family decision-tree classifiers.}
    \label{fig:drift_accdiff_corr_summary_2mff}
\end{figure}

\begin{figure}
    \centering
    \includegraphics[width=1\linewidth]{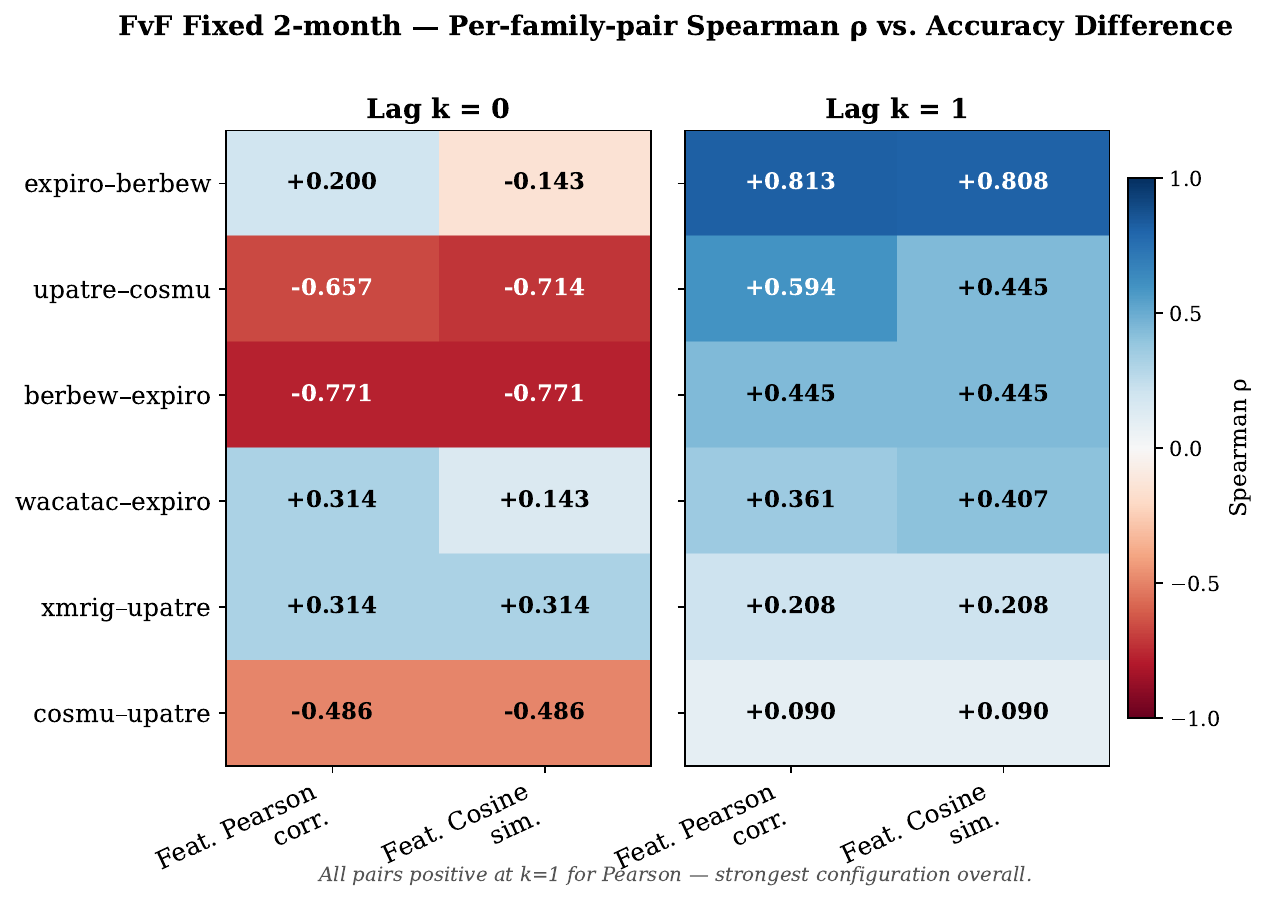}
    \caption{Spearman correlation between accuracy difference and drift metrics for family-versus-family decision-tree classifiers. Results are shown for lag $k=0$ (same window) and $k=1$ (drift preceding accuracy change).}
    \label{fig:famvfam_accdiffcorr_spearman_2m}
\end{figure}

The aggregated correlations between the drift metrics and the accuracy difference are summarized in  Figure \ref{fig:drift_accdiff_corr_summary_2mff}. The strongest correlations are again obtained using feature importance-based metrics. In this case, the first two correlations are stronger on average, reaching 0.4 and standard deviation being lower, indicating that per-family results may exhibit more stable results. Per-family results shown in Figure \ref{fig:famvfam_accdiffcorr_spearman_2m} indeed support this observation. The correlations are generally stronger and more consistent than in the family-versus-benign case, with peak values reaching above 0.8 at lag $k{=}1$. However, the magnitude of the correlations still varies between family pairs and drift metrics, typically ranging from approximately 0.2 to 0.8.

Overall, these results indicate that concept drift is more consistently captured in the family-versus-family setting than in the family-versus-benign setting. This suggests that drift signals may be easier to detect when the classifier is forced to discriminate between closely related malware families, where changes in family-specific behavior are less likely to be masked by the stable characteristics of benign samples.

\subsubsection{Family-versus-family data shift correlation}
Data shift correlations in the FvF setting are generally stronger than in the FvB case. The highest mean correlations reach approximately 0.35--0.39, obtained for Feature L1/L2 distance paired with Mean L2 data shift. Similarity-based feature importance metrics show negative correlations with data shift metrics, consistent with the expected inverse relationship. Per-family results remain family-dependent, with strong correlations for some families (e.g., wacatac reaching 0.8) and weak or opposite trends for others.

\subsection{Clustering}
We replaced fixed-time windowing with temporal segmentation using DBSCAN ($\text{eps\_days}{=}4$, $\text{min\_samples}{=}500$) and KMeans ($k{=}4$) to identify malware campaign periods. In the FvB setting, DBSCAN yielded peak correlations comparable to fixed windowing but with substantially higher variance and frequent sign reversals across families and lags. KMeans produced marginally weaker average correlations with characteristic $\pm 1.0$ saturation effects caused by extremely small sample sizes (3--4 transition pairs per family). In the FvF setting, both clustering algorithms produced weak mean correlations below 0.10 for accuracy difference, substantially lower than the fixed-windowing FvF result (mean $\rho = 0.419$). Data shift correlations in the DBSCAN FvF setting were moderate (strongest combination reaching $-0.398$), but remained family-dependent with no systematic advantage over fixed windowing. Overall, clustering-based window construction does not improve drift detection in any of the tested configurations.

\subsection{Summary of results across configurations}
For summarization we introduce two shortcuts, family versus family scenarios are noted as FvF and family versus benign scenarios as FvB.
\begin{table*}[htbp]
\centering
\caption{Summary of best aggregated Spearman correlations between drift metrics and accuracy difference across experimental configurations. Bold indicates the strongest result per configuration.}
\label{tab:summary_all_configs}
\begin{tabular}{llccc}
\toprule
\textbf{Configuration} & \textbf{Best drift metric} & \textbf{Mean $\rho$} & \textbf{Median $\rho$} & \textbf{Std.\ $\rho$} \\
\midrule
\multicolumn{5}{l}{\textit{Fixed windowing, accuracy correlation}} \\
FvB, 2-month       &  $\FeatureLOne$         & $-$0.224 & $-$0.216 & 0.389 \\
FvB, 2-week         & Jaccard cover rules         &    0.109 &    0.125 & 0.090 \\
\textbf{FvF, 2-month} & \textbf{Feature Pearson Correlation} & \textbf{0.419} & \textbf{0.403} & \textbf{0.262} \\
FvF, 2-month        & Feature Cosine Similarity   &    0.400 &    0.426 & 0.246 \\
\midrule
\multicolumn{5}{l}{\textit{Fixed windowing, data shift correlation}} \\
FvB, 2-month        & Feat.\ Cosine $\times$ KS mean &  0.247 &  0.288 & 0.367 \\
FvF, 2-month        & Feat.\ Cosine $\times$ Mean L2  & $-$0.391 & $-$0.353 & 0.461 \\
FvF, 2-month        & Feat.\ L1 $\times$ Mean L2      &  0.353 &  0.268 & 0.342 \\
\midrule
\multicolumn{5}{l}{\textit{Clustering, accuracy correlation}} \\
FvB, DBSCAN         & Activation stability drift  &  0.239 &  0.216 & 0.711 \\
FvB, KMeans         & Activation stability drift  &  0.161 &  0.491 & 0.932 \\
FvF, DBSCAN         & Feature L1 distance         &  0.038 &  0.005 & 0.342 \\
FvF, KMeans         & Coverage mean diff.\ windows &  0.087 &  0.042 & 0.198 \\
\midrule
\multicolumn{5}{l}{\textit{Clustering, data shift correlation}} \\
FvB, DBSCAN         & Feat.\ L1 $\times$ Domain\_Auc  &  0.339 &  0.607 & 0.678 \\
FvB, KMeans         & Pred.\ Agree.\ $\times$ KS mean &  0.619 &  0.522 & 0.283 \\
FvF, DBSCAN         & Feat.\ Pearson $\times$ Wasserstein & $-$0.398 & $-$0.417 & 0.262 \\
FvF, KMeans         & Pred.\ Agree.\ $\times$ Domain\_Auc & $-$0.266 & $-$0.304 & 0.219 \\
\bottomrule
\end{tabular}
\end{table*}

As illustrated in Table \ref{tab:summary_all_configs}, the fixed-windowing family-versus-family configuration consistently produces the strongest and most reliable results, with feature Pearson correlation achieving a mean $\rho$ of 0.419 and standard deviation of 0.262. Clustering-based approaches produce substantially weaker results in both the accuracy difference and data shift settings. The per-family results in Table \ref{tab:fvf_fixed_vs_clustering_perfamily} further illustrate this: whereas fixed-windowing FvF produces moderate positive correlations across most families, clustering yields mixed signs with no consistent pattern. The FvB KMeans results retain their characteristic $\pm$1.0 saturation due to extremely small sample sizes (3--4 transition pairs), which remains a fundamental limitation of clustering with few clusters.

\begin{table*}[htbp]
\centering
\caption{Per-family Spearman correlations ($k{=}1$) for the best drift metric in each configuration.}
\label{tab:fvf_fixed_vs_clustering_perfamily}
\begin{tabularx}{\textwidth}{l X cccccc} 
\toprule
\textbf{Configuration} & \textbf{Best metric} & \textbf{berb.} & \textbf{wac.} & \textbf{exp.} & \textbf{cos.} & \textbf{xmr.} & \textbf{upa.} \\
\midrule
\multicolumn{8}{l}{\textit{Accuracy difference correlation}} \\
FvF fixed 2m & Feature Pearson Correlation & 0.45 & 0.36 & 0.81 & 0.09 & 0.21 & 0.59 \\
FvF DBSCAN     & Feature L1 Distance & $-$0.18 & $-$0.40 & $-$0.13 & 0.55 & 0.25 & 0.14 \\
FvF KMeans     & Cov.\ mean diff & $-$0.02 & $-$0.15 & 0.00 & 0.08 & 0.42 & 0.18 \\
\midrule
\multicolumn{8}{l}{\textit{Data shift correlation}} \\
FvF fixed 2-mo & Feature L1 Distance $\times$ Mean L2  & $-$0.14 & 0.10 & 0.74 & $-$0.05 & 0.14 & 0.43 \\
FvF DBSCAN     & Feature Pearson Correlation $\times$ Wasserstein & $-$0.30 & $-$0.53 & $-$0.18 & $-$0.06 & 0.31 & $-$0.01 \\
\midrule
\multicolumn{8}{l}{\textit{FvB clustering}} \\
FvB KMeans acc & Activ.\ stability & $-$1.00 & 0.66 & 0.98 & 1.00 & $-$1.00 & 0.33 \\
FvB KMeans shift & Pred.\ agr.\ $\times$ KS mean & $-$1.00 & $-$1.00 & $-$1.00 & 1.00 & 1.00 & $-$1.00 \\
\bottomrule
\end{tabularx}
\end{table*}

\subsection{Comparison to other methods}
This subsection compares our best performing configuration (FvF fixed windowing with decision trees) against Transcendent \cite{barbero2022transcending} and the RIPPER rule-based classifier.

\subsubsection{Transcendent}
We evaluated Transcendent using both the ICE and CCE conformal evaluators in the same FvF fixed-windowing scenario. ICE produced correlations comparable to our method, with a peak of $0.854$ for the wacatac--expiro pair. However, neither method dominates across all pairs: Transcendent achieves stronger correlations where our method is weaker (e.g., wacatac--expiro: $0.854$ vs $0.361$; cosmu--upatre: $0.542$ vs $0.090$) and vice versa. The CCE evaluator performed substantially worse, producing zero or strongly negative correlations for most pairs, likely due to high variance when applied to shallow decision trees. These results suggest that Transcendent and our structural metrics capture complementary aspects of drift, sample-level distributional deviation versus decision boundary changes, and a practical monitoring system could benefit from combining both signals.

\subsubsection{RIPPER}

\begin{table*}[t]
\centering
\caption{Comparison of drift detection methods in the family-versus-family setting (2-month fixed windowing, Spearman correlation with accuracy difference).. All values correspond to lag $k{=}1$.}
\label{tab:comparison_methods}
\begin{tabular}{llccc}
\toprule
\textbf{Method} & \textbf{Drift signal} & \textbf{Mean $\rho$} & \textbf{Median $\rho$} & \textbf{Std.\ $\rho$} \\
\midrule
\textbf{Decision tree} & \textbf{Feature Pearson Correlation}     & \textbf{0.419} & \textbf{0.403} & \textbf{0.262} \\
\textbf{Decision tree} & \textbf{Feature Cosine Similarity} & \textbf{0.400} & \textbf{0.426} & \textbf{0.246} \\
RIPPER        &  Feature Pearson Correlation & 0.329 & 0.350 & 0.445 \\
RIPPER        & Feature Cosine Similarity & 0.288 & 0.251 & 0.423 \\
Transcendent (ICE) & rejection rate         & 0.230 & 0.149 & 0.420 \\
Transcendent (CCE) & rejection rate         & $-$0.079 & 0.000 & 0.662 \\
\bottomrule
\end{tabular}
\end{table*}

To test whether our metrics generalize beyond decision trees, we repeated the best-performing configuration using RIPPER. Feature importance-based metrics again yielded the strongest correlations, with feature Pearson correlation achieving a mean of $0.329$ (vs $0.419$ for decision trees). Per-family results revealed strong correlations for specific pairs (upatre--cosmu: $0.949$, expiro--berbew: $0.623$), but unlike decision trees, RIPPER also produced negative and zero correlations at $k{=}1$, explaining the higher standard deviation. Data shift correlations followed a comparable pattern, with mean values reaching approximately $0.2$. 
Overall, RIPPER confirms that our structural drift metrics generalize to other rule-based classifiers, though decision trees provide more consistent results, likely due to RIPPER's greedy sequential covering introducing greater sensitivity to individual windows. Table \ref{tab:comparison_methods} summarizes the comparison in the FvF 2-month setting: decision trees achieve the highest mean correlation (0.419) with the lowest standard deviation, followed by RIPPER ((.329) and Transcendent ICE (0.230). The higher variance of RIPPER and Transcendent reflects their complementary strengths on specific family pairs, reinforcing that combining structural and sample-level drift signals is a promising direction.

\section{\uppercase{Conclusions and Future Work}}
\label{sec:conclusion}
This work proposed structural drift metrics derived from decision tree rulesets and evaluated them across multiple configurations on the EMBER2024 dataset. By comparing symbolic rule representations across temporal windows, the approach provides both drift signals and interpretable insight into classifier changes.

The results demonstrate that decision tree rulesets evolve as malware families change over time, and that metrics derived from these ruleset changes can serve as meaningful indicators of concept drift. In particular, drift metrics based on feature importance and rule coverage consistently correlate with both accuracy degradation and data distribution shifts. However, no single drift metric exhibits uniform magnitude or direction of correlation across all families, indicating that drift behavior is inherently family-dependent.

Across all evaluated configurations, concept drift was found to be detectable but strongly influenced by temporal resolution, classifier formulation, and population selection. Drift signals were more stable and pronounced in family-versus-family classifiers than in family-versus-benign settings, where benign data often masked family-specific changes. Overall, the findings highlight the importance of interpretable, ruleset-based drift metrics and family-specific analysis for reliable monitoring of concept drift in malware detection systems.

Several directions for future work remain. First, the proposed metrics could be extended to deeper decision trees or ensemble methods such as random forests, where aggregating drift signals across multiple trees may improve robustness. Second, evaluating the proposed metrics on a simulated adversarial dataset with controlled drift injection would provide a more rigorous validation of their detection capability. Third, the evaluation could be expanded to cover all pairwise combinations of the six malware families and to additional datasets beyond EMBER2024 to assess generalizability. Finally, developing adaptive thresholding strategies that account for family-specific drift behavior would bring the approach closer to deployment in real-time malware detection systems.

\section*{\uppercase{Acknowledgements}}
This work was supported by the 2025 FIT CTU Student Summer Research Program in Prague and by the Grant Agency of the Czech Technical University in Prague, grant No. SGS26/187/OHK3/3T/18 funded by the MEYS of the Czech Republic.

\bibliographystyle{apalike}
{\small
\bibliography{example}}

\end{document}